\documentclass[reprint, amsmath,amssymb,pra]{revtex4-1}

\usepackage{graphicx}
\usepackage{dcolumn}
\usepackage{bm}
\usepackage{braket}
\usepackage{color}
\allowdisplaybreaks

\begin{document}

\abovedisplayskip=7pt
\abovedisplayshortskip=0pt
\belowdisplayskip=7pt
\belowdisplayshortskip=7pt

\preprint{APS/123-QED}

\title{Vacuum Induced Coherence in Cavity Quantum Electrodynamics}

\author{A. Vafafard$^{1}$}
\altaffiliation{azarvafafard@gmail.com}

\author{S. Hughes$^{2}$}
\author{G. S. Agarwal$^{3, 4}$}
\affiliation{$^{1}$Department of Physics, University of Zanjan, University Blvd, 45371-38791, Zanjan, Iran}
\affiliation{$^{2}$Department of Physics, Queen's University, Kingston, Ontario, Canada, K7L 3N6}
\affiliation{$^{3}$Institute for Quantum Science and Engineering and Department of Biological and Agricultural Engineering,
Texas A\&M University, College Station, Texas 77845, USA}
\affiliation{$^{4}$Department of Physics, Oklahoma State University, Stillwater, Oklahoma 74078, USA}
\date{\today }

\begin{abstract}
Vacuum induced coherence in a strongly coupled cavity consisting of a three-level system is studied theoretically. The effects of the strong coupling to electromagnetic field vacuum are examined by solution of an open-system quantum master equation. The numerical results show that the system exhibits population trapping, and the numerical results are interpreted with analytical expressions derived from a new  basis in the weak excitation regime. We further show that the generated effects can be probed with weak external fields. Moreover, it is shown that the induced coherence can be controlled by the applied field parameters like field detuning.     Finally, we  study the trapping dynamics in the strong field excitation regime, and also demonstrate that a recently proposed asymmetric pumping regime (limited to the weak coupling regime) can  remove the radiative decay of coherent Rabi oscillations, with both weak and strong excitation fields.
\end{abstract}


\maketitle

\section{\label{sec:level1}Introduction}

It is now well established that quantum coherence is an underlying principle for controlling the optical properties of a  medium. Quantum coherence induced by the interaction of  coherent laser fields with a quantum system has a key role in many implementations in optical physics. In recent years, a new class of quantum coherence generation, focusing on coherence induced by the vacuum of  electromagnetic fields, has been the subject of extensive studies both experimentally and theoretically. 
The study of vacuum induced coherency (VIC) has been triggered by Agarwal \cite{Agarwal} and continues to be of much interest \cite{VIC}. It has been demonstrated that, in the absence of any applied field, quantum interference occurs between the decay channels and leads to a coherence based incoherent process \cite{Ficek}. 
It is worth noting that spontaneous emission, which is a source of decoherence  in quantum processes, is a major source of difficulty in many quantum optical phenomena such as single photon emission. However, it has been shown that spontaneous emission reduction~\cite{reduction1,reduction2} and cancellation~\cite{cancellation1,cancellation2,cancellation3} are possible  by exploiting VIC.  Indeed,
the manifestation of VIC leads to numerous studies like modified resonance fluorescence~\cite{RF1,RF2}, fluorescence quenching \cite{FQ1,FQ2}, unexpected population inversion \cite{PI}, gain with or without inversion \cite{gain}, phase dependent line shape \cite{phase}, ultracold photoassociative ro-vibrational excitations \cite{ultracold},  optical generation of electron spin coherence
in quantum dots \cite{QDs}, and long-lived quasistationary coherences \cite{Quasistationary Coherences}. 

The deep connection between quantum mechanics and thermodynamics is also a current base for some investigations of novel heat engines. The consistency of dynamical equations with thermodynamic equilibrium has been shown for the interaction between a heat bath and a multilevel atom \cite{heat bath}. The quantum dynamics of a V-type system driven by weak coupling to a thermal bath, relevant to light harvesting processes, has also been studied \cite{Vtype}. Recently, VIC is predicted as a new way to increase the power of a quantum heat engines (QHE) which transform high-energy thermal radiation into low entropy useful work. Scully {\it et al.} showed that it is possible to break detailed balance and enhance quantum efficiency in a QHE without using any external field and energy source \cite{QH}. The power enhancement of heat engines in a degenerate V-type three level system has also been studied \cite{power enhancement}. 
Another motivation in this area is to estimate the role of VIC on biological QHE, where it is demonstrated that the photosynthetic reaction center may be considered as the biological QHE that converts hot thermal radiation into electron flux. Interestingly, it has been shown that VIC is the common origin of population oscillations in photosynthetic complexes and enhancement of photo current \cite{QHC}. Recently, Hughes and Agarwal also showed how
semiconductor  cavity systems can induce population trapping
and substantial anisotropic  AVI between orthogonal dipole states in single quantum dots~\cite{HughesAndAgarwal}.

To the best of our knowledge,  all the above studies considered weak coupling to the vacuum field which allows the standard Born approximation in the calculations. In this study, on the contrary, we consider a strong coupling to the vacuum and analyze the appearance of the VIC effects. Then, we present a model that involves an atom (or any three level quantum system such as a quantum dot) inside a cavity. The population trapping is a considerable achievement of coherency in the presented model. 
Our paper is organized as follows. In Sec. \ref{equations}, the atomic model is presented. In Sec. \ref{results}, we analyze the results of the system in three parts. In the first part, \ref{sec:III.1}, the behavior of the system in the vacuum of electromagnetic field is studied. We investigate the time evolution of population that shows a trapping of population in upper states occurs. By defining a new set of states and deriving analytical expressions, the population trapping is interpreted theoretically. In the second part,  \ref{sec:III.2}, the generated effects are probed by applying two external fields. These fields are sufficiently weak that  we use a weak excitation approximation (WEA).
In the third part,  \ref{sec:III.3}, we look at the breakdown
of the WEA with stronger excitation fields, and also study
a double pumping scenario ($\Omega_\alpha=-\Omega_\beta$),
and confirm the absence of any radiative decay of coherent Rabi oscillations---as recently predicted in the weak cavity coupling regime \cite{HughesAndAgarwal}. We conclude and present closing discussions in Sec.~\ref{Conclusions}.

\section{\label{equations}Theoretical MODEL AND EQUATIONS}

\subsection{Energy level system and master equations}
\begin{figure}
\centering
  \includegraphics[clip,trim=0cm 0cm 0cm 0.cm,width=0.99\columnwidth]{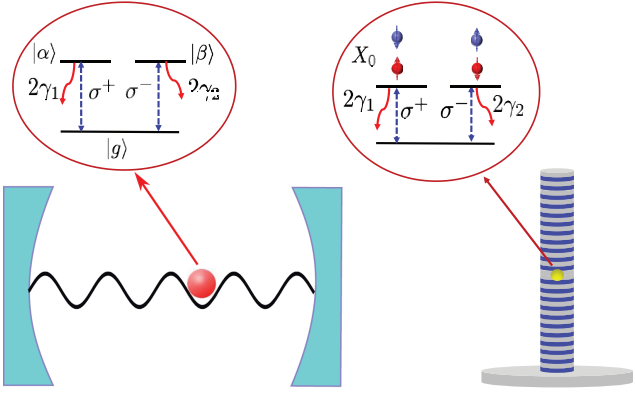}
\caption{\small (left) A simple three level atom system trapped in a leaky optical cavity with the decay constant $ \kappa $. (right) Three-level quantum dot system of a single charge-neutral exciton in a semiconductor cavity (e.g., pillar microcavity).  }
\label{fig1}
 \end{figure}

We consider a single three-level atom which, e.g., is trapped in a leaky cavity with the decay constant, $ \kappa $. An example atomic level scheme is shown in Fig.~\ref{fig1} (left). The cavity mode is linearly polarized through $(\sigma^{+}+\sigma^{-})/ \sqrt{2} $. For example, the dipole-allowed transitions $\left|g\right\rangle\equiv \left [ F=4, M_{F}=-4 \right]$ and $\left|\alpha\right\rangle\equiv \left [ F=5, M_{F}=-3 \right] $ ( $\left|\beta\right\rangle\equiv \left [F=5, M_{F}=-5 \right] $) are coupled by the $ \sigma^{+} $ ($ \sigma^{-} $) polarized component of the cavity mode. Here $ F $ is the total atomic angular momentum quantum number and $ M_{F} $ represents the magnetic quantum number of the corresponding state. The spontaneous decay rates from levels $ \left|\alpha\right\rangle $ and $\left|\beta\right\rangle$ to $\left|g\right\rangle$ are denoted by $ 2\gamma_{1} $ and  $ 2\gamma_{2} $, respectively. Such a system can be generated in the $ D_{2} $ line of $ ^{133}Cs $ atom \cite{model}, or in  semiconductor quantum dots~\cite{QD8,QD9,QD10,QD11,HughesAndAgarwal} (see Fig.~\ref{fig1} (right)). 

The quantized cavity field at the atom can be  expressed as
\begin{equation}
\hat{\bf E}=a_{0}\hat{\epsilon}_{x}  a  e^{-i\omega_{c} t}+ {\rm H.c.},
\end{equation}
where {H.c.} refers to Hermitian conjugate, $a_{0}=i\sqrt{2\pi\hbar \omega_{c}/V}$ (in Gaussian units, with $V$ is the effective mode volume), and $\omega_{c} $ is the cavity mode frequency. The dipole operator is
\begin{equation}
\hat{\bf d}=\left|d\right|\hat{\epsilon}_{+}e^{-i\omega_{0} t}\left|g\right\rangle\langle\alpha|+\left|d\right|\hat{\epsilon}_{-}e^{-i\omega_{0} t}\left|g\right\rangle\langle\beta|+{\rm H.c.},
\end{equation}
where $ \omega_{0} $ denotes the atomic transition frequency. 
The system Hamiltonian, in the dipole and rotating wave approximations, is given by (rotating frame at $\omega_c$):
\begin{align}
H_{S} & =
-\hbar \Delta\left|\alpha\right\rangle\langle\alpha|
-\hbar\Delta \left|\beta\right\rangle\langle\beta|\nonumber \\
&-\hbar g (\left|g\right\rangle\langle\alpha|a^{\dagger}+\left|g\right\rangle\langle\beta|a^{\dagger})+{\rm H.c.}.
\end{align}
where we have introduced the atom-cavity detuning $\Delta=\omega_0-\omega_c$,
and the dipole-cavity coupling rate
$g=\sqrt{\pi\omega_{c}/V\hbar}$. 

In the presence of a coherent driving field, we can drive the atoms
directly with some Rabi field $\Omega$, and obtain 
the following system Hamiltonian (now in a frame rotating at the 
laser frequency $\omega_L$),
\begin{align}
H_{\rm S}/\hbar & =
{\Omega_\alpha} ({\sigma}_{g\alpha}+ {\sigma}_{\alpha g})
+ \Omega_\beta ({\sigma}_{g\beta}+ {\sigma}_{\beta g}) \nonumber \\
&+a^\dagger a(\omega_c-\omega_L)+ \sigma_{\alpha\alpha}(\omega_0-\omega_L)
+ \sigma_{\beta\beta} (\omega_0-\omega_L)\nonumber \\
&- g (\left|g\right\rangle\langle\alpha|a^{\dagger}
+\left|g\right\rangle\langle\beta|a^{\dagger})+{\rm H.c.}.
\end{align}

The quantum master equation for the density matrix operator $ \rho $ of the system can then be written as 
\begin{align}
\frac{\partial\rho}{\partial t} &= \frac{-i}{\hbar}[H_S,\rho]-\kappa(a^{\dag}a\rho-2a\rho a^{\dag}+\rho a^{\dag}a)\nonumber\\
& -\gamma_{1}(A_{\alpha g}\rho A_{g\alpha}-2A_{g\alpha}\rho A_{\alpha g}+\rho A_{\alpha g}A_{g\alpha})\nonumber\\
& -\gamma_{2}(A_{\beta g}\rho A_{g\beta}-2A_{g\beta}\rho A_{\beta g}+\rho A_{\beta g}A_{g\beta}) ,
\label{eq:ME}
\end{align}
where $ A_{ij}$ is an operator  defined through $ A_{ij}=|i\rangle \langle j| , (i,j=\alpha, \beta, g)$.

 \subsection{Weak excitation approximation}
In the limit of weak excitation (here, in the limit of one quantum excitation) or vacuum
dynamics only, the four basic states for this system are defined as
follows:
\begin{align}
|\psi_{1}\rangle=|\alpha\rangle|0\rangle,~~~~~~~~~~~~
|\psi_{2}\rangle=|\beta\rangle|0\rangle,\nonumber\\
|\psi_{3}\rangle=|g\rangle|1\rangle,~~~~~~~~~~~~~
|\psi_{4}\rangle=|g\rangle|0\rangle.
\end{align}
The corresponding density matrix equations of motion  then take the following form: 
\begin{subequations}
\begin{align}
\dot{\rho}_{\psi 1\psi 1} &= i g \rho_{\psi 3\psi 1} - i g \rho_{\psi 1\psi 3} - 2 \gamma_{1} \rho_{\psi 1\psi 1},
\\ 
\dot{\rho}_{\psi 2\psi 2}  &=  i g \rho_{\psi 3\psi 2} - i g \rho_{\psi 2\psi 3} - 2 \gamma_{2} \rho_{\psi 2\psi 2}, \\
\dot{\rho}_{\psi 3\psi 3}  &= i g \rho_{\psi 1\psi 3} - i g \rho_{\psi 3\psi 1} +i g \rho_{\psi 2\psi 3}- i g \rho_{\psi 3\psi 2}\nonumber\\
&  - 2 \kappa \rho_{\psi 3\psi 3},
\\
\dot{\rho}_{\psi 4\psi 4}  &= 2 \gamma_{1} \rho_{\psi 1\psi 1}+2 \gamma_{2} \rho_{\psi 2\psi 2}+2 \kappa \rho_{\psi 3\psi 3}, \\
\dot{\rho}_{\psi 1\psi 2}   &= i g \rho_{\psi 3\psi 2} - i g \rho_{\psi 1\psi 3} - ( \gamma_{1} +\gamma_{2})\rho_{\psi 1\psi 2},\\
\dot{\rho}_{\psi 1\psi 3}   &= - i \Delta \rho_{\psi 1\psi 3}+ i g (\rho_{\psi 3\psi 3}-\rho_{\psi 1\psi 1}) 
\nonumber 
\\
& {}-i g \rho_{\psi 1\psi 2} -( \gamma_{1} +\kappa) \rho_{\psi 1\psi 3}, \\
\dot{\rho}_{\psi 1\psi 4}   &=  i g \rho_{\psi 3\psi 4}  -\gamma_{1} \rho_{\psi 1\psi 4},
\\
\dot{\rho}_{\psi 2\psi 3}   &= - i \Delta \rho_{\psi 2\psi 3}+ i g (\rho_{\psi 3\psi 3}-\rho_{\psi 2\psi 2}) \nonumber\\
& {}- i g \rho_{\psi 2\psi 1} - \gamma_{2} \rho_{\psi 2\psi 3}-\kappa \rho_{\psi 2\psi 3},
\\
\dot{\rho}_{\psi 2\psi 4}   &=  i g \rho_{\psi 3\psi 4}  -\gamma_{2} \rho_{\psi 2\psi 4},
\\
\dot{\rho}_{\psi 3\psi 4}   &= i \Delta \rho_{\psi 3\psi 4}+ i g \rho_{\psi 1\psi 4} + i g \rho_{\psi 2\psi 4}\nonumber\\
&  -\kappa \rho_{\psi 3\psi 4},
\end{align}
\label{DME}
\end{subequations}

\section{\label{results}Results and Discussions }
Here we present the main results. First, in  \ref{sec:III.1}, we  study the behavior of the system in the absence of external fields. Second, in \ref{sec:III.2}, we examine vacuum induced properties by using a weak external field, in the limit of weak excitation. 
Third, in \ref{sec:III.3}, we explore the regime of high field pumping
and solve the master equation in a basis of cavity photon states.
We also study a regime of asymmetric field (Rabi) pumping.

\subsection{\label{sec:III.1}VACUUM INDUCED COHERENCE}

\begin{figure}[b]
 \centering
   \includegraphics[clip,trim=0cm 0cm 0cm 0.cm,width=1\columnwidth]{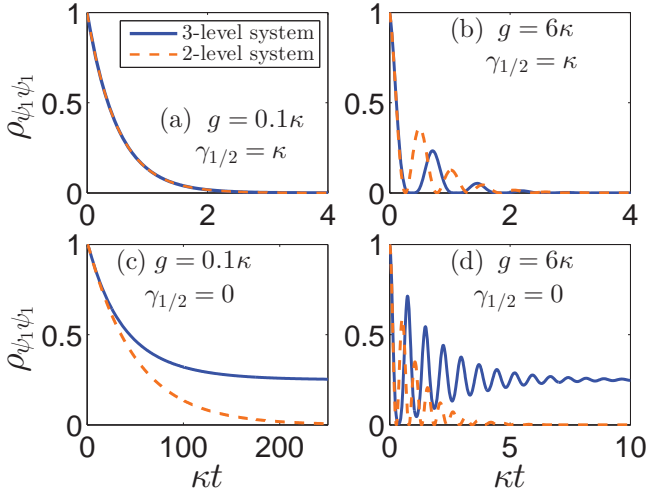}
\caption{\small Time evolution of $ \rho_{\psi 1\psi 1} $  in the presence (a,b) and absence (c,d) of spontaneous decay rates for different values of the coupling constant. The decay time is shown in units of the cavity decay constant, $\kappa$. The solid (dashed) lines show the 3-level (2-level) results.}
\label{fig2}
 \end{figure}

We first investigate the dynamical behavior of the system by solving equation
set (\ref{DME}). It is assumed that the atom is prepared initially in the state $ \left|\psi_{1}\right\rangle $, so  $ \rho_{\psi 1\psi 1}(t=0)= 1$. All other parameters  are normalized with respect to $ \kappa$ (half the cavity decay rate). Figure \ref{fig2} shows the population $ \rho_{\psi 1\psi 1} $, for nonzero (a, b) and zero (c, d) values of spontaneous decay rates. This plot is depicted for different values of the cavity-atom coupling constant: $g=0.1k$ (a, c), $6k$ (b, d). The solid curve describes the behavior of the  three level system, while the dashed shows the results of a two level system (i.e., without interference effects). For $ \gamma=0 $, the solid curve exhibits a nonzero value for time, indicating population trapping in the upper state. To help elucidate the physical origin of such trapping, we compare this situation with the case in which only the state $\left|\alpha\right\rangle $  is coupled to $\left|g\right\rangle$ (dashed curve  in Fig. \ref{fig2}). Comparing the two curves shows that the existence of the second channel for light decay is responsible for population trapping.

\begin{figure}
\centering
   \includegraphics[width=0.75\columnwidth]{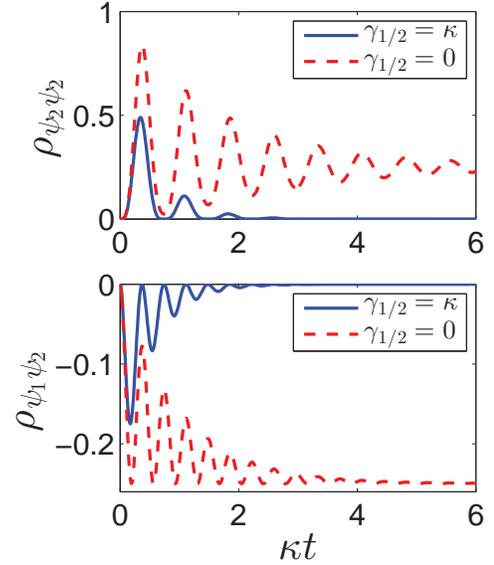}
\caption{\small Time evolution of $ \rho_{\psi 2\psi 2} $ (a)
and $ \rho_{\psi 1\psi 2} $ (b). The parameters are $g=6\kappa $ and $ \Delta=0 $.}
\label{fig3}
 \end{figure}

To better understand the  trapping phenomena through the electromagnetic field vacuum, we can make a transformation to a new basis defined by  
\begin{align}
\left|\varphi_{1}\right\rangle   &= \frac{\left|\psi_{1}\right\rangle+\left|\psi_{2}\right\rangle}{\sqrt{2}},
\left|\varphi_{2}\right\rangle = \frac{\left|\psi_{1}\right\rangle-\left|\psi_{2}\right\rangle}{\sqrt{2}},
\nonumber \\
\left|\varphi_{3}\right\rangle   &= \left|\psi_{3}\right\rangle,
\left|\varphi_{4}\right\rangle = \left|\psi_{4}\right\rangle, 
\label{new basis}
\end{align}
where
in this new basis, the state $ \left|\varphi_{2}\right\rangle$ does not couple to the cavity field, and is thus termed a ``dark state''. Consequently, with spontaneous emission, there is no population decays to the lower state. We will see that the analytical results, derived by the new basis equation, are fully confirmed by the numerical solutions. We start our calculations with the density matrix equations in this new basis: 
\begin{subequations}
\begin{eqnarray}
\dot{\rho}_{\varphi_{1}\varphi_{1}} & = &- i \sqrt{2} g \rho_{\varphi_{1}\varphi_{3}}+i \sqrt{2} g \rho_{\varphi_{3}\varphi_{1}} - 2 \gamma \rho_{\varphi_{1}\varphi_{1}}, 
\\
\dot{\rho}_{\varphi_{2}\varphi_{2}} & = & - 2 \gamma \rho_{\varphi_{2}\varphi_{2}}, 
\\
\dot{\rho}_{\varphi_{3}\varphi_{3}} & = & i \sqrt{2} g \rho_{\varphi_{1}\varphi_{3}}-i \sqrt{2} g \rho_{\varphi_{3}\varphi_{1}} - 2 \kappa\rho_{\varphi_{3}\varphi_{3}},
\\
\dot{\rho}_{\varphi_{4}\varphi_{4}} & = &  2 \gamma( \rho_{\varphi_{1}\varphi_{1}}+ \rho_{\varphi_{2}\varphi_{2}})+2 \kappa\rho_{\varphi_{3}\varphi_{3}}, 
\\
\dot{\rho}_{\varphi_{1}\varphi_{2}} & = & i \sqrt{2} g \rho_{\varphi_{3}\varphi_{2}}- 2 \gamma \rho_{\varphi_{1}\varphi_{2}}, 
\\
\dot{\rho}_{\varphi_{1}\varphi_{3}} & = &  i \sqrt{2} g (\rho_{\varphi_{3}\varphi_{3}} -\rho_{\varphi_{1}\varphi_{1}}) - \gamma \rho_{\varphi_{1}\varphi_{3}} - \kappa\rho_{\varphi_{1}\varphi_{3}}, 
\\
\dot{\rho}_{\varphi_{1}\varphi_{4}} & = & i \sqrt{2} g \rho_{\varphi_{3}\varphi_{4}} - \gamma \rho_{\varphi_{1}\varphi_{4}}, 
\\
\dot{\rho}_{\varphi_{2}\varphi_{3}} & = &- i \sqrt{2} g \rho_{\varphi_{2}\varphi_{1}} - \gamma \rho_{\varphi_{2}\varphi_{3}} - \kappa\rho_{\varphi_{2}\varphi_{3}}, 
\\
\dot{\rho}_{\varphi_{2}\varphi_{4}} & = & - \gamma \rho_{\varphi_{2}\varphi_{4}}, 
\\
\dot{\rho}_{\varphi_{3}\varphi_{4}} & = & i \sqrt{2} g \rho_{\varphi_{1}\varphi_{4}} -  \kappa\rho_{\varphi_{3}\varphi_{4}}, 
\label{new density matrix}
\end{eqnarray} 
\end{subequations}
where we have set $ \gamma_{1}= \gamma_{2}= \gamma $. 
Using the new basis, an analytical expression for $ \rho_{\psi{1}\psi{1}}$ can be found, 
\begin{eqnarray}
 \rho_{\psi{1}\psi{1}}(t) &=& \dfrac{-1}{4B^{2}}e^{-2 t \gamma}[-B^{2}-4 g^{2} e^{t(\gamma-\kappa)}\nonumber\\
 && {} +2 M B e^{ t(\gamma -\kappa)/2}+N e^{ t(\gamma -\kappa)}],
 \label{rho11}
\end{eqnarray}
where
\begin{eqnarray}
B & = &\sqrt{8g^{2}-(\gamma-\kappa)^{2}} ,\nonumber\\
M& = &-B \cos(Bt/2)+(\gamma-\kappa)\sin(Bt/2) ,\nonumber\\
N& = &(-4g^{2}+(\gamma-\kappa)^{2})\cos(Bt)+B(\gamma-\kappa)\sin(Bt)).\nonumber
\end{eqnarray}

An inspection of Eq.~(\ref{rho11}) reveals that the population is trapped in upper states for the case of $ \gamma=0 $. The VIC is achieved here by quantum interference of two channels, coupled with the cavity field, for no spontaneous decay rates. 
We can simplify the Eq.~(\ref{rho11}) by assuming $ \gamma=0 $,  yielding
\begin{equation}
 \rho_{\psi{1}\psi{1}}(t)=-\dfrac{-B^{2}-4g^{2}e^{- t \kappa}+2 B  e^{- t \kappa /2}(m)+e^{  -t \kappa}(n)}{4B^{2}},
  \label{rhoN11}
\end{equation}   
where $ m=M(\gamma=0) $ and $ n=N(\gamma=0) $.
Equation (\ref{rhoN11}) shows that, in long time limit,  $ \rho_{\psi{1}\psi{1}}(t\rightarrow\infty) = 1/4$.

Figure \ref{fig3}(a-b) displays the population of $\left|\psi_{2}\right\rangle$ and $ \rho_{\psi 1\psi 2} $, respectively. The main parameters are $ g=6\kappa$, and $ \Delta=0$. The same behavior can be seen in these figures, namely $ \rho_{\psi{2}\psi{2}}(t\rightarrow\infty) = 1/4$ and $ \rho_{\psi{1}\psi{2}}(t\rightarrow\infty) =- 1/4$.
 
\subsection{\label{sec:III.2}PROBING THE VACUUM INDUCED COHERENCE BY EXTERNAL WEAK FIELDS}

Now the population trapping and coherence created can be studied by using two external weak fields (WEA). The probe field with Rabi frequency $G_{1} (G_{2})$ is applied to the transition $ |\alpha\rangle-|g\rangle $ ($ |\beta\rangle-|g\rangle $). The density matrix equations in the presence of external fields take the form:
\begin{subequations}
\label{FDMEset}
\begin{align}
\dot{\rho}_{\psi 1\psi 1} & =  i g \rho_{\psi 3\psi 1} - i g \rho_{\psi 1\psi 3} +i G_{1} \rho_{\psi 4\psi 1}-i G_{1} \rho_{\psi 1\psi 4} 
\nonumber 
\\
&{} - 2 \gamma_{1} \rho_{\psi 1\psi 1},
\\
\dot{\rho}_{\psi 2\psi 2} & =  i g \rho_{\psi 3\psi 2} - i g \rho_{\psi 2\psi 3}+i G_{2} \rho_{\psi 4\psi 2}-i G_{2} \rho_{\psi 2\psi 4} 
\nonumber 
\\
&{} - 2 \gamma_{2} \rho_{\psi 2\psi2},
\\
\dot{\rho}_{\psi 3\psi 3} & =  i g \rho_{\psi 1\psi 3} - i g \rho_{\psi 3\psi 1} +i g \rho_{\psi 2\psi 3} - i g \rho_{\psi 3\psi 2} 
\nonumber 
\\
&{}- 2 \kappa \rho_{\psi 3\psi 3},
\\
\dot{\rho}_{\psi 4\psi 4} & =  i G_{1} (\rho_{\psi 1\psi 4}- \rho_{\psi 4\psi 1})+i G_{2} (\rho_{\psi 2\psi 4}-\rho_{\psi 4\psi 2})\nonumber \\
&{} +2 \gamma_{1} \rho_{\psi 1\psi 1}+2 \gamma_{2} \rho_{\psi 2\psi 2}+2 \kappa \rho_{\psi 3\psi 3},
\\
\dot{\rho}_{\psi 1\psi 2} & =  i g \rho_{\psi 3\psi 2} - i g \rho_{\psi 1\psi 3}+i G_{1} \rho_{\psi 4\psi 2}-i G_{2} \rho_{\psi 1\psi 4}  
\nonumber 
\\
&{}- ( \gamma_{1} +\gamma_{2})\rho_{\psi 1\psi 2},
\\
\dot{\rho}_{\psi 1\psi 3} & =   i g (\rho_{\psi 3\psi 3}-\rho_{\psi 1\psi 1}) - i g \rho_{\psi 1\psi 2}+i G_{1} \rho_{\psi 4\psi 3} \nonumber \\
 &{}- \gamma_{1} \rho_{\psi 1\psi 3}-\kappa \rho_{\psi 1\psi 3},
 \\
\dot{\rho}_{\psi 1\psi 4} & =  i \delta \rho_{\psi 1\psi 4}+i g \rho_{\psi 3\psi 4}  +i G_{1}  (\rho_{\psi 4\psi 4}-\rho_{\psi 1\psi 1}) \nonumber \\
&{} -\gamma_{1} \rho_{\psi 1\psi 4},
\\
\dot{\rho}_{\psi 2\psi 3} & =   i g (\rho_{\psi 3\psi 3}-\rho_{\psi 2\psi 2}) - i g \rho_{\psi 2\psi 1} +i G_{2} \rho_{\psi 4\psi 3} \nonumber \\
&{}- \gamma_{2} \rho_{\psi 2\psi 3}-\kappa \rho_{\psi 2\psi 3},
\\
\dot{\rho}_{\psi 2\psi 4} & =  i \delta \rho_{\psi 2\psi 4}-i G_{1} \rho_{\psi 2\psi 1}+i G_{2}  (\rho_{\psi 4\psi 4}-\rho_{\psi 2\psi 2}) 
\nonumber 
\\  
 &{}+i g \rho_{\psi 3\psi 4}-\gamma_{2} \rho_{\psi 2\psi 4},
\\
\dot{\rho}_{\psi 3\psi 4} & =   i \delta \rho_{\psi 3\psi 4}-i G_{1} \rho_{\psi 3\psi 1}-i G_{2} \rho_{\psi 3\psi 2}+ i g \rho_{\psi 1\psi 4}
\nonumber 
\\  
&{} + i g \rho_{\psi 2\psi 4} -\kappa \rho_{\psi 3\psi 4},
\label{FDME}
\end{align}
\end{subequations}
where $ \delta $ denotes the probe field detuning with the atomic resonance transition. 

\begin{figure}
\centering
   \includegraphics[width=1\columnwidth]{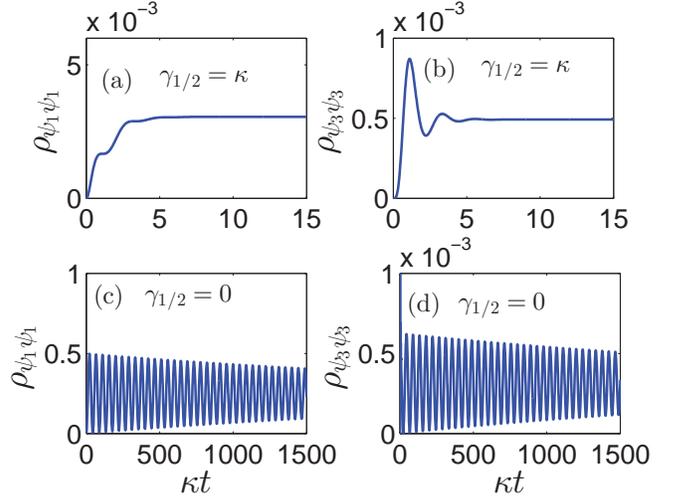}
\caption{\small Time evolution of $\rho_{\psi 1\psi 1}$ (a,c) and $\rho_{\psi 3\psi 3}$ (b,d). The parameters are $g=2\kappa, G_{1}=0.1 \kappa, G_{2}=0 $, and $ \Delta=\delta=0$ Time evolution of $ \rho_{\psi 1\psi 2} $. }
\label{fig4}
 \end{figure}

\begin{figure}[h]
\centering
   \includegraphics[width=0.75\columnwidth]{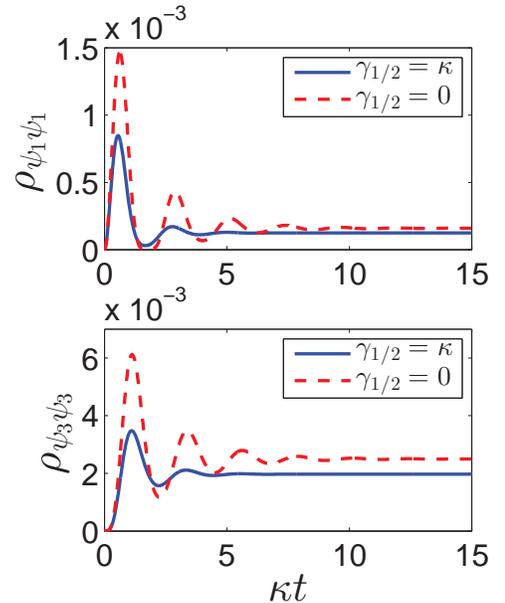} 
\caption{\small Time evolution of $ \rho_{\psi 1\psi 1}$ (a) and $ \rho_{\psi 3\psi 3} $ (b). The parameters are $g=2\kappa, G_{1}= G_{2}=0.1 \kappa $, and $ \Delta=\delta=0 $.}
\label{fig5}
 \end{figure}

Next, we  present the numerical and analytical results under various parametric conditions. We solve time dependent equation set (\ref{FDMEset}) numerically. First, we consider $ G_{2}=0 $. The time evolution of $\rho_{\psi 1\psi 1}$ for $\gamma_1=\gamma_2=\kappa$ (a) and $\gamma_1=\gamma_2=0$ (c) is shown  in Fig.~\ref{fig4}. As we expect with zero  spontaneous  emission decay, the population is trapped in the upper states and this trapping can be easily seen by tracing the probe field. We also find that applying only one external field leads to a very slow approach to steady state. This is because of the field induced small decay. To investigate this investigation, we calculate the cavity output field which is proportional to  $ \langle a^{\dagger}a\rangle $. By using the following states for this system, we can obtain the mean cavity photon number: 
\begin{align}
\langle a^{\dagger} a\rangle & = \sum_{i, j}\langle\psi_{j}\vert a^{\dagger} a\vert\psi_{i}\rangle \rho_{\psi_{j}\psi_{i}},\nonumber\\
 & = \langle\psi_{3}\vert a^{\dagger} a\vert\psi_{3}\rangle \rho_{\psi_{3}\psi_{3}},\nonumber\\
  & = \langle 0\vert\langle g\vert g\rangle\vert 0\rangle \rho_{\psi_{3}\psi_{3}},\nonumber\\
 & = \rho_{\psi_{3}\psi_{3}},
\end{align}
where we have used the relation $ a\vert\psi_{i}\rangle=\langle\psi_{i}\vert a^{\dagger} =0, (i=1, 2, 4)$. The time evolution of $ \rho_{\psi 3\psi 3} $ is shown by Fig. \ref{fig4}(b). The output field is a good confirmation for population trapping in upper states. Also, the time dependent output can be seen for the case $ \gamma_{1}=\gamma_{2}=0 $.  In the next step, we switch on both of the probe fields and investigate the results for different conditions.  
Figure \ref{fig5}(a) shows the population of the upper states. The solid and dashed curves display the result for $ \gamma_{1}=\gamma_{2}=\kappa $  and $ \gamma_{1}=\gamma_{2}=0 $ , respectively. The other parameters are $g=2\kappa$,  $ G_{1}= G_{2}=0.1 \kappa$, and $ \Delta=\delta=0 $. These conditions are also considered for plot of $ \rho_{\psi 3\psi 3} $ in Fig.~\ref{fig5}(b).

  
 To derive the analytical expression in the presence of external applied fields, we can again use the new basis density matrix equations which can be rewritten as 
 \begin{subequations}
 \label{ndmSet}
 \begin{align}
\dot{\rho}_{\varphi_{1}\varphi_{1}} & = - i \sqrt{2} g (\rho_{\varphi_{1}\varphi_{3}}-i  \rho_{\varphi_{3}\varphi_{1}})- i \sqrt{2} G (\rho_{\varphi_{1}\varphi_{4}}-\rho_{\varphi_{4}\varphi_{1}})\nonumber\\
 &{} - 2 \gamma \rho_{\varphi_{1}\varphi_{1}}, 
 \\
\dot{\rho}_{\varphi_{2}\varphi_{2}} & =  - 2 \gamma \rho_{\varphi_{2}\varphi_{2}}, 
\\
\dot{\rho}_{\varphi_{3}\varphi_{3}} & =  i \sqrt{2} g \rho_{\varphi_{1}\varphi_{3}}-i \sqrt{2} g \rho_{\varphi_{3}\varphi_{1}} - 2 \kappa\rho_{\varphi_{3}\varphi_{3}},
\\
\dot{\rho}_{\varphi_{4}\varphi_{4}} & =  i \sqrt{2} G (\rho_{\varphi_{1}\varphi_{4}}-i  \rho_{\varphi_{4}\varphi_{1}}) +2 \gamma( \rho_{\varphi_{1}\varphi_{1}}+ \rho_{\varphi_{2}\varphi_{2}})
\nonumber
\\
&{}+2 \kappa\rho_{\varphi_{3}\varphi_{3}}, 
\\
\dot{\rho}_{\varphi_{1}\varphi_{2}} & =  i \sqrt{2} g \rho_{\varphi_{3}\varphi_{2}}+ i \sqrt{2} G \rho_{\varphi_{4}\varphi_{2}}- 2 \gamma \rho_{\varphi_{1}\varphi_{2}}, 
\\
\dot{\rho}_{\varphi_{1}\varphi_{3}} & =   i \sqrt{2} g (\rho_{\varphi_{3}\varphi_{3}} -\rho_{\varphi_{1}\varphi_{1}})+ i \sqrt{2} G \rho_{\varphi_{4}\varphi_{3}} - \gamma \rho_{\varphi_{1}\varphi_{3}}\nonumber
\\
&{} - \kappa\rho_{\varphi_{1}\varphi_{3}}, 
\\
\dot{\rho}_{\varphi_{1}\varphi_{4}} & =   i \delta \rho_{\varphi_{1}\varphi_{4}} + i \sqrt{2} G (\rho_{\varphi_{4}\varphi_{4}} -\rho_{\varphi_{1}\varphi_{1}})+i \sqrt{2} g \rho_{\varphi_{3}\varphi_{4}}
\nonumber
\\
&{} - \gamma \rho_{\varphi_{1}\varphi_{4}}, 
\\
\dot{\rho}_{\varphi_{2}\varphi_{3}} & = - i \sqrt{2} g \rho_{\varphi_{2}\varphi_{1}} - \gamma \rho_{\varphi_{2}\varphi_{3}} - \kappa\rho_{\varphi_{2}\varphi_{3}}, 
\\
\dot{\rho}_{\varphi_{2}\varphi_{4}} & =  i \delta \rho_{\varphi_{2}\varphi_{4}}-i \sqrt{2} g \rho_{\varphi_{1}\varphi_{2}} - \gamma \rho_{\varphi_{2}\varphi_{4}}, 
\\
\dot{\rho}_{\varphi_{3}\varphi_{4}} & =  i \delta \rho_{\varphi_{3}\varphi_{4}}+i \sqrt{2} g \rho_{\varphi_{1}\varphi_{4}}-i\sqrt{2} G \rho_{\varphi_{3}\varphi_{1}}  -  \kappa\rho_{\varphi_{3}\varphi_{4}}.\nonumber\\ 
\label{ndm}
\end{align}
\end{subequations}

\begin{figure}
\centering
    \includegraphics[width=0.75\columnwidth]{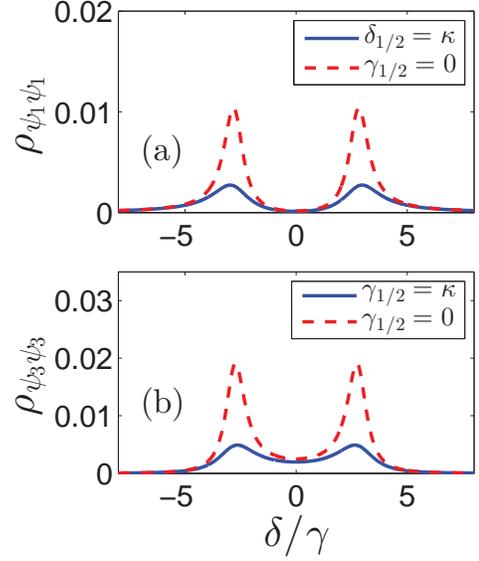}
\caption{\small The $ \rho_{\psi 1\psi 1} $ (a) and $ \rho_{\psi 3\psi 3} $ (b) versus probe field detuning. The parameters are $g=2\kappa, G_{1}= G_{2}=0.1 \kappa,$  and  $ \Delta=0 $. }
\label{fig6}
 \end{figure}

 
As we are interested to study the steady-state result of the system, in Fig. \ref{fig6}(a) we show the $ \rho_{\psi _{1}\psi_{1}} $ versus probe field detuning in the steady-state condition. This behavior can be confirmed by analytical results, and solving equation set (\ref{ndmSet}) in steady-state leads to 
\begin{equation}
\rho_{\psi _{1}\psi_{1}}=\rho_{\psi _{2}\psi_{2}}=\rho_{\psi _{1}\psi_{2}}=\dfrac{G^{2}(\delta^{2}+\kappa^{2})}{A},
\end{equation}
\begin{equation}
\rho_{\psi _{3}\psi_{3}}=\dfrac{4G^{2}g^{2}}{A},
\end{equation}
where $ A=(4g^{4}-4g^{2}\delta^{2}+\gamma^{2} \delta^{2}+4g^{2}\gamma\kappa+\gamma^{2}\kappa^{2}+\delta^{4}+\delta^{2}\kappa^{2}) $.
The maximum values of $ \rho_{\psi 1\psi 1} $ appear in $ \delta=\pm2.73\kappa $. The vacuum induced trapping is clearly seen for these values of the field detunings. 

In Fig. \ref{fig6}(b), we plot $ \rho_{\psi 3\psi 3} $ versus probe field detuning. This figure is another evidence of population trapping for $ \gamma=0 $. It is also obvious that a larger field output is achieved for the case in which the atom is trapped.  

\subsection{\label{sec:III.3}High Field Regime}

Having analyzed the vacuum and weak field regimes, we now study the high field regime, which requires a full numerical solution
to the master equation (\ref{eq:ME}), with the required number
of cavity photon states. We are thus in the multi-phonon anharmonic cavity-QED\ regime.

\begin{figure}[t!]
\centering
   \includegraphics[width=0.75\columnwidth]{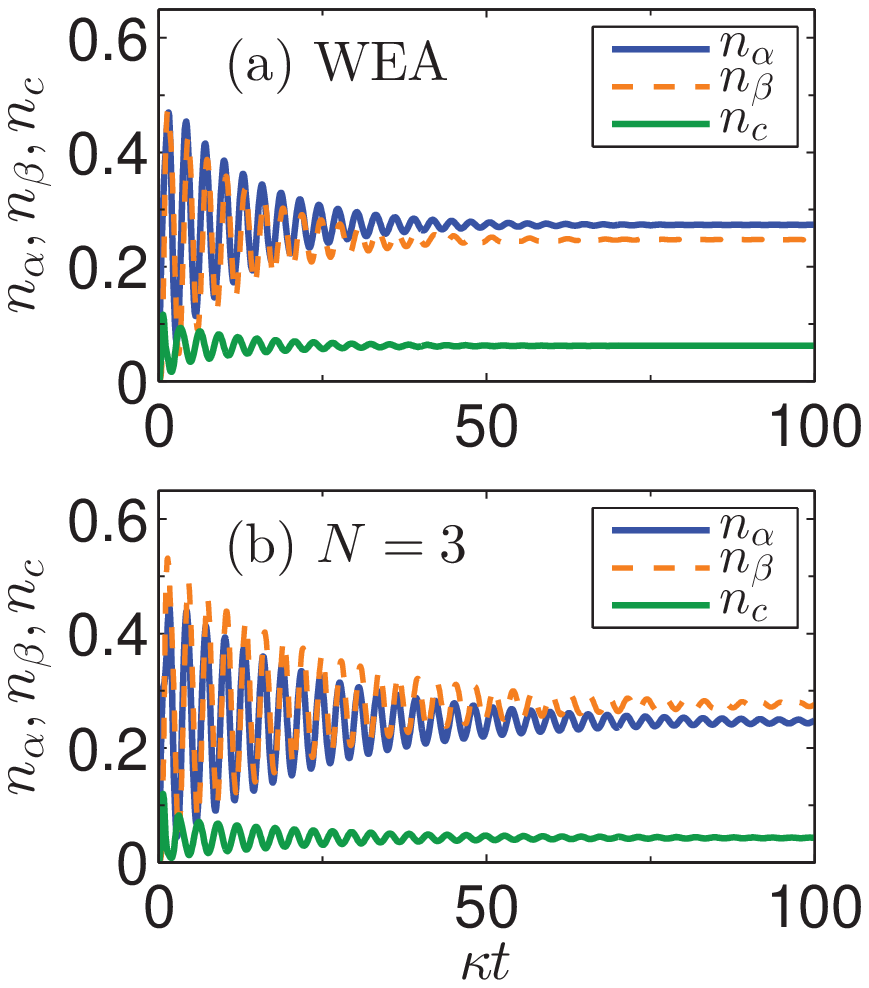}
\caption{\small Time evolution of the populations
$n_\alpha,n_\beta,n_c$ when $\Omega_\alpha=\kappa$,
using the WEA ($N=2$, i.e., two cavity states) (a) and multi-photon regime with
$N=3$ cavity states (b). The cavity-atom coupling rate
is $g=2\kappa$ and $\gamma=0$, so $\Omega_\alpha=g/2$.}
\label{fig7}
 %
 %
\vspace{0.5cm}
   \includegraphics[width=0.75\columnwidth]{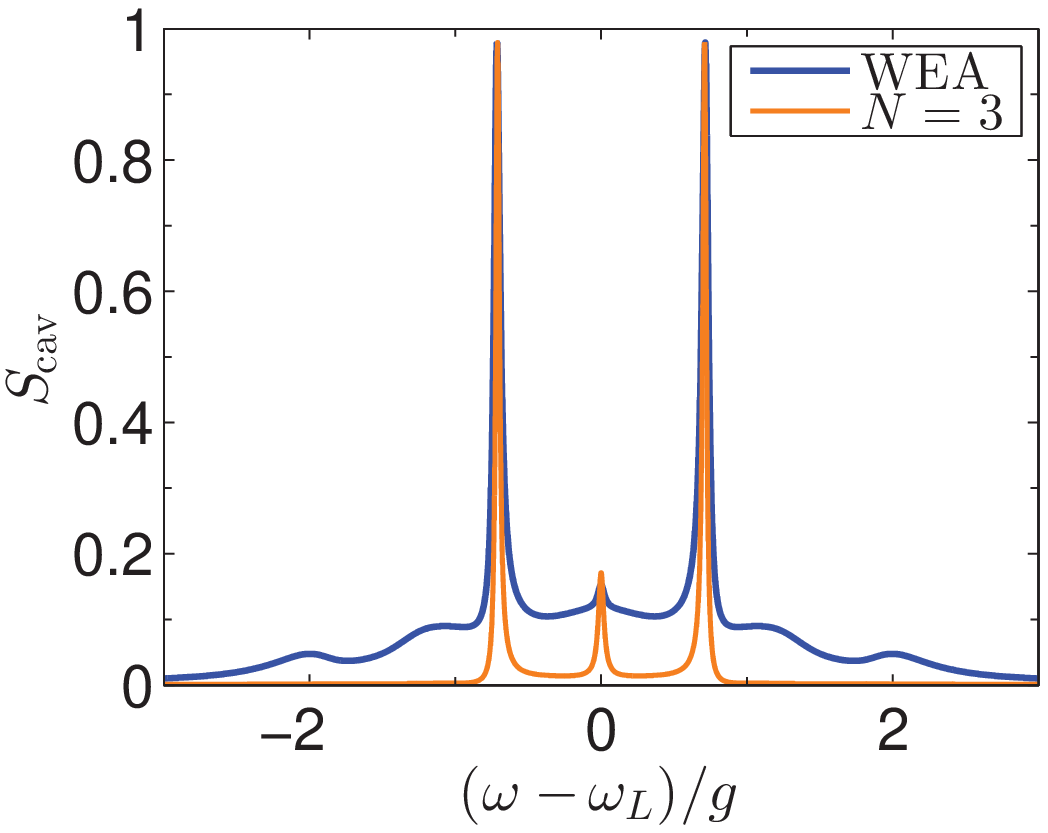}
\caption{\small Cavity-emitted spectra corresponding to 
the excitation regimes in Fig.~\ref{fig7}.}
\label{fig7b}
 \end{figure}

We again consider the case of $g=2\kappa$,
and choose on- resonance pumping with $\Delta=0=\delta$,
and $\gamma=0$.  Under these conditions, we find that
the WEA works  well up for pump strengths
up to $\Omega \approx 0.5\kappa$. The WEA is defined here
as using a maximum of $N=2$ cavity photon states
in the numerical solution of (\ref{eq:ME}).

 In Figs.~\ref{fig7}(a)-(b),
we show the population dynamics of the two
excited state levels ($n_\alpha=\braket{A_{\alpha \alpha}}$,$n_\beta=\braket{A_{\alpha \alpha}}$) and the cavity photon number ($n_c=\braket{a^\dagger a}$), with and without the WEA, where we clearly start to see differences in the predicted
populations. In particular, we find that the population damping has a longer decay with 
multiphoton processes, and $n_\beta> n_\alpha$ in the long time limit. In addition, the long time mean cavity photon numbers
are overestimated by around 40\% within the WEA.
 For the multiphoton calculations, a basis
of $N=3$ cavity photon states was found to be sufficient to ensure accurate numerical convergence.
To help understand this WEA breakdown, and the longer decay dynamics
of the multi-photon result,
 we 
 compute
the cavity-emitted spectrum:
\begin{align}
S_{\rm cav}(\omega)& = \lim_{t\rightarrow\infty}\text{Re}[\int_0^{\infty}d\tau(\braket{a^\dagger(t+\tau)a(t)} \nonumber \\
&-\braket{a^\dagger(t)}\braket{a(t)})e^{i(\omega_L-\omega)\tau}],
\end{align}
where we employ the quantum regression theorem
to obtain the two-time correlation function. Figure \ref{fig7b}
demonstrates that the WEA significantly overestimates the spectral width for the resonances,
and obtains the incorrect oscillator strength of some of the dressed-state resonances. For the WEA displayed here, which uses three atom states and $N=2$ cavity states,
the dressed-state eigenenergies with $\Omega_\alpha=g/2$ are 
$E_i/g= 1.5388, 0.3633, 0,  0,  -0.3633, -1.5388$, which are the quasienergies, separated from the next photon manifold by $\omega_L$. In contrast, for the $N=3$ basis,
then we have
$E_i/g=2.1167,    1.4644,    0.3539,   0,   0,    0,   -0.3539,   -1.4644,  -2.1167,$ further demonstrating that the WEA fails in this high excitation regime.

 \begin{figure}
\centering
   \includegraphics[width=0.75\columnwidth]{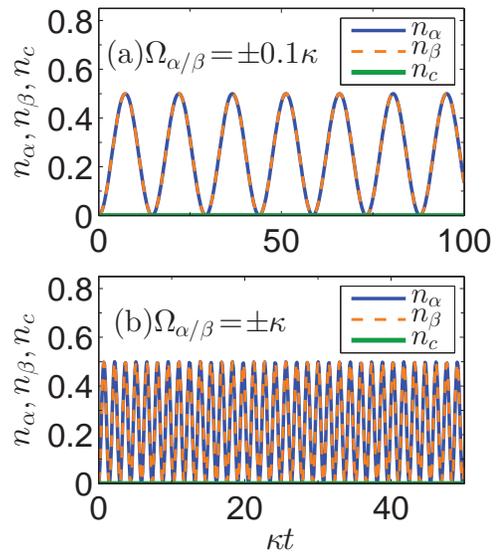}
\caption{\small 
Time evolution of the populations
$n_\alpha,n_\beta,n_c$ when $\Omega_\alpha=0.1\kappa=-\Omega_\beta$ (a)
and $\Omega_\alpha=\kappa=-\Omega_\beta$ (b). We solve the full
master equation with 
$N=3$ states, and as in Fig.~\ref{fig7} use   $g=2\kappa$ and $\gamma=0$.}
\label{fig8}
 \end{figure}

For our final study, we look at an antisymmetric double pumping regime
with $\Omega_\alpha=-\Omega_\beta$. This excitation regime is motivated by recent work that, within
the Born-Markov approximation for the weak coupling regime,
demonstrated that such an excitation regime can possibly eliminate
the radiative damping of field-driven coherent Rabi oscillations~\cite{HughesAndAgarwal}.
It is thus of interest to see if such features also occur in the strong coupling regime of cavity-QED. To investigate this situation, for both weak and strong excitation, we again solve the full master equation (\ref{eq:ME}). Using the
same system parameters as in Fig.~\ref{fig7}, we consider
the two different pump strength of
$\Omega_\alpha=-\Omega_\beta=0.1\kappa$
and $\Omega_\alpha=-\Omega_\beta=\kappa$,
which are shown, respectively, in Fig.~\ref{fig8}(a)
and Fig.~\ref{fig8}(b). In agreement with
the results in \cite{HughesAndAgarwal}, we find no damping of the 
coherent Rabi oscillations for either excitation regime and thus perfect Rabi oscillation of the population trapped states. In addition,
the cavity photon coupling remains dark, with no cavity population 
appearing.

\section{\label{Conclusions}CONCLUSIONS}
We have studied various VIC effects of an atom (or three-level system such as a quantum dot)\ located inside a cavity in the regime of cavity-QED. In our numerical results, we found that the population is trapped in the upper state. Using a new basis definition enables us to find analytical results that are useful to interpret the major physical mechanism of the population trapping. The quantum interference between two channels, which are coupled by cavity field, causes the trapping of the population in the upper states. Two weak probe fields, using a WEA, are applied to the system to demonstrate the VIC. We found that the results in the presence of only one external field depend on the time. By studying the external field intensity, the the population trapping is further confirmed. Moreover, the detuning of applied probe field was used as a controlling parameter. With suffiently syrong fields, we also showed that
multi-photon states can cause longer decay of the coherent Rabi oscillations, in a regime where the WEA fails.
Finally,
we also that an antisymmetric pumping regime ($\Omega_\alpha=-\Omega_\beta$)
can completely eliminate the decay of coherent Rabi oscillations in the cavity-QED\ regime, for both weak and strong excitation fields. 
\acknowledgements
This work was funded by the Biophotonics initiative of Texas A\&M
University, and the 
 Natural Sciences and Engineering Research Council of Canada.

\end{document}